\documentclass[apj]{emulateapj}
\usepackage{apjfonts}

\input epsf.tex

\begin{document}
\title{Calibration of photometry from the Gemini Multi-Object Spectrograph on Gemini North}
\author{Inger J{\o}rgensen}
\affil{Gemini Observatory}
\affil{670 N.\ A`ohoku Pl., Hilo, HI 96720, USA}
\email{ijorgensen@gemini.edu}

\submitted{Accepted for publication in PASA, November 4, 2008}

\begin{abstract}
All available observations of photometric standard stars
obtained with the Gemini Multi-Object Spectrograph at Gemini North in the
period from August 2001 to December 2003 have been used to establish the calibrations
for photometry obtained with the instrument.
The calibrations presented in this paper are based on significantly more
photometric standard star observations than usually used by the individual users.
Nightly photometric zero points as well as color terms are determined.
The color terms are expected to be valid for all observations taken prior to UT 2004 November 21
at which time the Gemini North primary mirror was coated with silver instead of aluminum.
While the nightly zero points are accurate to 0.02 mag or better (random errors), the
accuracy of the calibrations is limited by systematic errors from so-called ``sky concentration'',
an effect seen in all focal reducer instruments.
We conclude that an accuracy of 0.035 to 0.05 mag can be achieved by using
calibrations derived in this paper.
The color terms are strongest for very red objects, e.g.\ for objects with $(r'-z')=3.0$
the resulting $z'$ magnitudes will be $\approx$ 0.35 mag too bright if the color term is ignored.
The calibrations are of importance to the large Gemini user community
with data obtained prior to UT 2004 November 21, as well as future users of
achive data from this period in time.
\end{abstract}


\section{Introduction}

The Gemini Multi-Object Spectrograph (GMOS-N) at the Gemini North 8-meter telescope
was commissioned in August and September 2001, and has 
been in regular operation since October 2001.
GMOS-N is used in imaging mode for imaging of fields later to be observed
in multi-object spectroscopic mode as well as for imaging programs 
aimed at photometry of faint targets.
GMOS-N is equipped with a filter set ($u'\,g'\,r'\,i'\,z'$)
intended to be identical to the Sloan-Digital-Sky-Survey (SDSS) filters. 
The SDSS $u'\,g'\,r'\,i'\,z'$ standard star system is
described in Smith et al.\ (2002).
Filters designed to reproduce the SDSS system have also been used at the
Isaac Newton Telescope, in particular for the wide field survey carried out
with this telescope, see Lewis et al.\ (2000) for details.

Photometric standard star fields were observed with GMOS-N during each 
photometric night when imaging data were obtained for commissioning, system 
verification and/or queue programs.
Using all of these observations together offer the possibility of determining the photometric
calibration to an accuracy not possible from the photometric standard stars usually 
made available to individual investigators. Long-term monitoring of the system 
performance (telescope plus instrument) is possible, as well as testing for 
systematic effects in the calibrations.
Since all Gemini data are archived, the calibrations derived in this paper
will also be of importance to future archive users.

This paper presents the photometric zero points for
all photometric nights during which photometric standard stars were observed,
an accurate determination of the color terms, and
tests for systematic effects in the calibrations.
The intent is to calibrate the GMOS-N photometric system to consistency with the SDSS system. 
As for the SDSS system, the calibrated magnitudes are very close to AB magnitudes.
GMOS-N and the commissioning of the instrument are described in detail in Hook et al.\ (2004).
GMOS-S on Gemini South is similar to GMOS-N. However, the calibrations and color terms
are expected to be different for the two instruments primarily due to differences in
the detector quantum efficiencies.
See Ryder, Murrowood \& Stathakis (2006) for examples of photometric calibrations for GMOS-S.

The paper is organized as follows. Section 2 describes the 
available data, the observed fields, the basic reductions, and how the photometry is derived.
Section 3 compares the GMOS-N system throughput with the SDSS system
and describes the transformation of the Landolt (1992) standard magnitudes to the
SDSS system. In Section 4
we derive the nightly zero points and discuss the variations of these with time.
The color terms in the transformations are established in Section 5,
while Section 6 presents other tests for systematic effects.
The results are summarized in Section 7.

\section{Observational data, basic reductions and photometry \label{sec-obs}}

\subsection{Standard star fields and observations}

We have used all observations of photometric standard
star fields obtained with GMOS-N between UT 2001 August 20
and UT 2003 December 26. The photometric standard star
fields were chosen from Landolt (1992), except 
the field GJ745A+B observed in November 2003 which is from Smith et al.\ (2002).
While in retrospect it might have been better to select SDSS standard stars
from Smith et al.\ as standard stars for GMOS-N, no information was publicly
available on the SDSS magnitudes of these stars at the time when GMOS-N started operations.

Table \ref{tab-gmosstdstars} lists the names of the fields and the central coordinates.
There are eleven standard star fields normally used for GMOS-N, listed as
``standard field'' in Table \ref{tab-gmosstdstars}.
In addition the table contains 15 fields that were only observed during
November 2003. These fields contain red standard stars, most of them with
$(r'-z')>1.2$. They were observed in order to establish the color terms
for especially the $i'$-filter and the $z'$-filter in the far red.
These fields are listed as ``red star'' in Table \ref{tab-gmosstdstars}.
During November 2003, the two standard star
fields PG0231+051 and SA98-670 were observed multiple times.

\begin{deluxetable}{lrrl}
\tablecaption{GMOS-N Standard Star Fields\label{tab-gmosstdstars} }
\tablewidth{0pt}
\tablehead{
\colhead{Field name} & \colhead{R.A. (J2000)} & \colhead{Decl. (J2000)} & \colhead{Notes} 
}
\startdata
              SA92-250 &  0 54 43.35 & +0 41 15.5  & Standard field \\
            PG0231+051 &  2 33 38.00 & +5 18 33.0  & Standard field \\
              SA95-100 &  3 53 03.91 & +0 01 14.9  & Standard field \\
              SA98-670 &  6 52 08.94 & -0 20 54.2  & Standard field \\
                 RU149 &  7 24 16.00 & -0 32 45.9  & Standard field \\
            PG0942-029 &  9 45 12.14 & -3 07 52.3  & Standard field \\
             SA101-330 &  9 56 17.20 & -0 27 35.0  & Standard field \\
            PG1047+003 & 10 50 06.00 & -0 01 00.0  & Standard field \\
            PG1323-086 & 13 25 49.35 & -8 49 45.1  & Standard field \\
             SA110-361 & 18 42 47.35 & +0 08 05.0  & Standard field \\
            PG2213-006 & 22 16 24.00 & -0 21 27.0  & Standard field \\
                G97\_42 &  5 28 00.15 & +9 38 38.1  & Red star \\
               G102\_22 &  5 42 09.27 & +12 29 21.6  & Red star \\
                97\_345 &  5 57 33.18 & +0 21 16.5  & Red star \\
                 97\_75 &  5 57 55.08 & -0 09 28.5  & Red star \\
                98\_618 &  6 51 49.58 & -0 21 15.8  & Red star \\
           101\_268+262 &  9 56 13.20 & -0 31 03.1  & Red star \\
                G44\_27 & 10 36 01.27 & +5 07 11.1  & Red star \\
                G163\_6 & 10 42 54.16 & +2 47 20.6  & Red star \\
                G44\_40 & 10 50 52.06 & +6 48 29.3  & Red star \\
                G45\_20 & 10 56 29.05 & +7 00 52.2  & Red star \\
            G163\_51+50 & 11 08 03.03 & -5 11 38.6  & Red star \\
                G12\_43 & 12 33 17.27 & +9 01 16.7  & Red star \\
               GJ745A+B & 19 07 09.85 & +20 52 59.9  & Red star \\
                 G26\_7 & 21 31 18.61 & -9 47 26.4  & Red star \\
               G156\_31 & 22 38 32.33 &-15 18 16.3  & Red star
\enddata
\tablecomments{Units of right ascension are hours, minutes, and seconds, and
units of declination are degrees, arcminutes, and arcseconds.}
\end{deluxetable}

The imaging field of view of GMOS-N is approximately 5.5\,arcmin by 5.5\,arcmin.
The normal standard star fields, as well as the field 98\_618
contain between three and ten standard stars.
The other fields observed in November 2003 contain only one or two stars. 

Observations were done during 117 nights. These nights were either commissioning
nights or queue science nights. Usually the staff observer only obtained
standard star observations if the night was judged to be photometric based on
the counts from the guide-probes on Gemini North and/or the information from
the Canada-France-Hawaii-Telescope (CFHT) SkyProbe
(http://www.cfht.hawaii.edu/Instruments/Elixir/skyprobe/\\
tonight.html).
The photometric quality of the nights were checked after the fact as well
using archived data from these sources.
Data taken during non-photometric nights were excluded from the analysis.
A total of 1694 images were obtained during photometric nights, 282 of those
during the November 2003 observations.
Ninety different stars were observed, 66 of these are in the normal standard star
fields, while the remainder were only observed during November 2003.
Thirteen of the stars observed are included in Smith et al.\ (2002) as primary
standards for the SDSS system, six of these were only observed during November 2003.
Observations were done in $g'$, $r'$, $i'$, and $z'$ though not all filters
were used each night.
On seven nights also $u'$ observations were obtained.

The typical exposure times were 1 or 3 seconds for $g'$, $r'$, $i'$ and $z'$,
and 20 sec for $u'$. The GMOS-N shutter is a blade shutter. The travel time
for the shutter blades depends slightly on the direction of movement,
causing $\approx$ 1 percent uncertainty on the exposure time of 1 sec exposures.
Using the average of two exposures taken immediately after each other
instead of a single exposure eliminates this uncertainty. 
Thus, all the standard star observations were taken as pairs of two exposures.

\subsection{Basic reductions}

A custom program was written to automatically select the photometric standard star 
observations from all observations obtained with GMOS-N, and then select 
the mean bias image and twilight flat field closest in time to each standard star observation.
The images were then processed using the GMOS tasks in the Gemini IRAF package; v1.5 
of the package was used. 
The Gemini IRAF package is an external IRAF\footnote{IRAF is distributed by National 
Optical Astronomy Observatories, which
is operated by the Association of Universities for Research in Astronomy, Inc., (AURA),
under cooperative agreement with the National Science Foundation, USA.} 
package distributed by Gemini Observatory.
The images were bias subtracted, converted from 
ADU to electrons, and flat field corrected. Bias images and flat fields 
were the mean images derived for each dark-period
and are identical to those reduced calibration images distributed to the 
users with programs observed during queue operation of the instrument. 
Finally the images from the three detectors were mosaiced together to one image. 

Using the world-coordinate-system in the images together with improved coordinates 
for the Landolt standard stars provided by John Thorstensen (private communication, 2003), 
pixel coordinate files for all images were produced automatically.
The coordinates from Thorstensen were derived using the USNO A2.0
catalog (Monet et al.\ 1996) as well as the Carlsberg Meridian Circle CMC12 catalog
(Evans, Irwin \& Helmer 2002).
The CMC12 positions have external accuracies of $\sim 40$ mas and
an epoch near 2000; the USNO A2.0 positions are generally
accurate to $\sim 300$ mas and are from the POSS-I, taken in the
1950s.  Known high-proper-motion stars not in CMC12 were updated
manually to epoch 2000.
A few Landolt stars missing in Thorstensen's list, were added manually
using the USNO A2.0 catalog.
For other users, the improved coordinates are distributed as part
of Gemini IRAF package (v1.5 and later) in the table {\tt gmos\$calib/nlandolt.fits}.

The reduced images were processed with the task {\tt gemseeing},
which is part of the Gemini IRAF package, in order to 
determine the full-width-half-maximum (FWHM) and ellipticity of the 
point-spread-function (PSF).
Only the standard stars in each field were used for this purpose. 
Saturated stars were eliminated from the measurements. 
Images with very poor image quality were omitted from further analysis. 
The poor image quality is due to the fact that majority of the standard star 
observations were done without guiding. When observing without guiding,
the Gemini Telescopes are subject to wind shake, which on occasion can cause 
very poor image quality even in exposures with exposure time below 5 seconds.

\begin{figure*}
\epsfxsize=16.5cm
\epsfbox{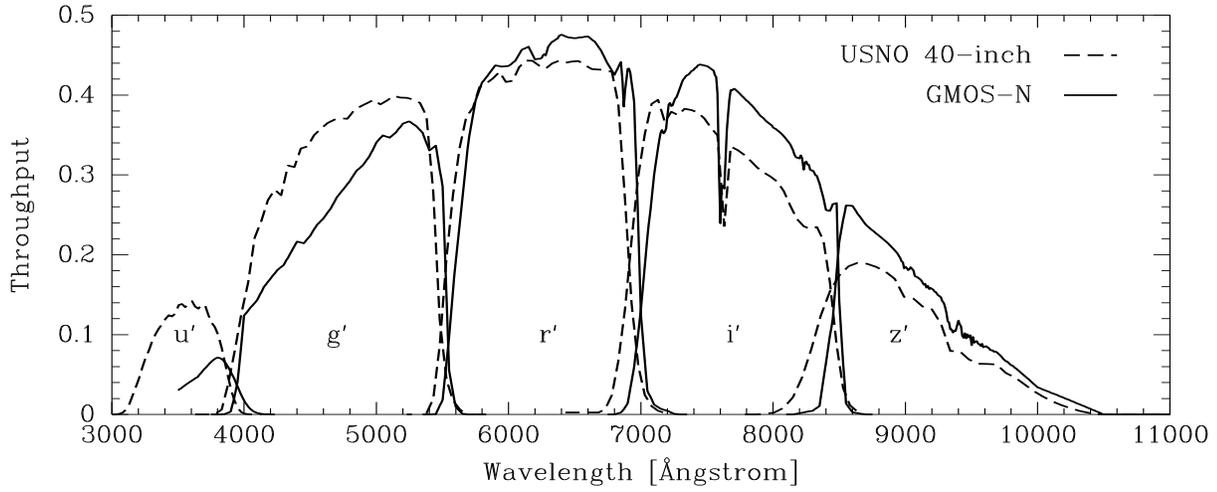} 
\caption[]{System throughput in the five SDSS filters. Solid lines --  GMOS-N on Gemini North,
including the telescope and the instrument (camera optics, filters and CCDs). 
Dashed lines -- USNO 40-in telescope, including the telescope, filters and CCD. 
In both cases, the atmospheric extinction at the site is included for airmass=1.0.
\label{fig-throughput}}
\end{figure*}

\begin{figure*}
\epsfxsize=16.5cm
\epsfbox{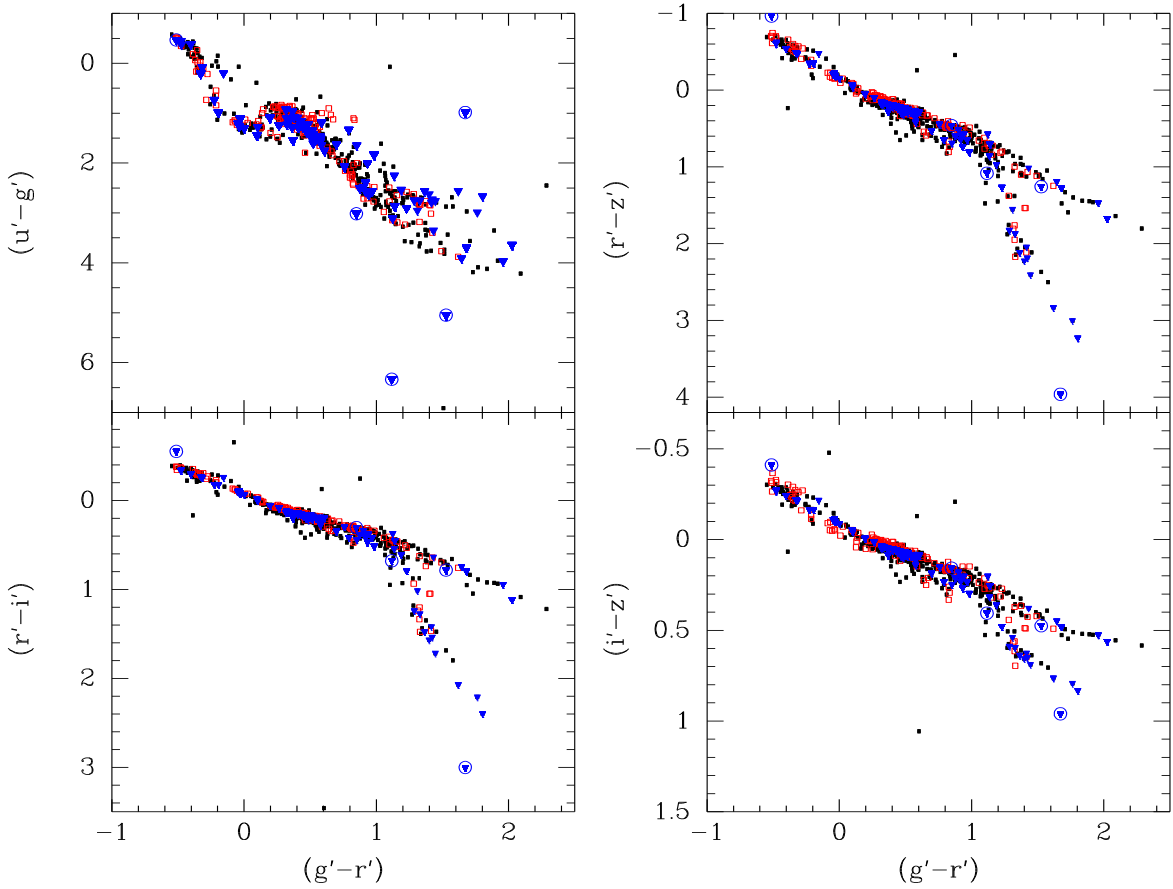}
\caption[]{Color-color diagrams of the standard stars.
Small solid boxes (black) -- Landolt (1992) transformed to the SDSS system;
open boxes (red) -- primary standard stars for the SDSS system from Smith et al.\ (2002);
triangles (blue) -- the stars observed with GMOS-N, magnitudes are from Landolt (1992) transformed to the SDSS system;
triangles in circles (blue) -- the stars observed with GMOS-N that were excluded from the 
determination of the nightly magnitude zero points and also excluded from some of the fits,
see text for details. \label{fig-colorcolor}}
\end{figure*}

\subsection{Photometry \label{sec-phot}}

For each image the median values of the FWHM and the ellipticity, $\epsilon$,
were used to determine the aperture size to be used for the photometry.
The aperture radius for each image was derived as
\begin{equation}
r_{\rm aperture}= 4 \cdot {\rm FWHM} \sqrt{(1+\epsilon )/(1-\epsilon )}
\end{equation}
For a Moffat PSF with $\beta = 2.8$, as is typical for GMOS-N images,
less than 0.7 per cent of the signal is expected to be outside this aperture size.
Thus, no aperture corrections were applied to the photometry.
See Trujillo et al.\ (2001) for a graphical comparison of Gaussian and Moffat
profiles. In the limit of $\beta = \infty$ the Moffat profile is identical to a
Gaussian profile.

Instrumental magnitudes were then derived as 
\begin{equation}
m_{\rm inst} = 28 - 2.5 \log (N/t) - k\,(airmass-1)
\end{equation}
where $t$ is the exposure time, $N$ is the number of electrons inside the aperture 
and above the sky level, $airmass$ is the mean airmass for the exposure, 
and $k$ is the median atmospheric extinction at Mauna Kea. 
The adopted median atmospheric extinction is 
0.42, 0.14, 0.11, 0.10 and 0.05 for $u'$, $g'$, $r'$, $i'$, and $z'$, respectively
(B\'{e}land, Boulade \& Davidge 1988).

\section{The SDSS photometric system \label{sec-calib}}

\subsection{Throughput for Gemini North with GMOS-N and for the USNO 40-inch telescope}

The SDSS primary standards were observed with the USNO 40-inch telescope,
Flagstaff, Arizona (Smith et al.\ 2002). 
In this section we compare the available throughput
information for that system with the throughput of the Gemini North 8-meter telescope
plus GMOS-N.
Figure \ref{fig-throughput} shows the system throughput for the two systems.
The data for the USNO 40-inch are from the web site
http://home.fnal.gov/$\sim$dtucker/ugriz/Filters/response.html,
while the data for Gemini are available on the GMOS-N web pages (linked from http://www.gemini.edu).
The transmission curves for the filters used in GMOS-N appear to be shifted slightly redwards relative
to those of the filters used at the UNSO 40-inch. Further, GMOS-N has relatively
better throughput in the red and poorer throughput in the blue, compared to the 
USNO 40-inch. This is most likely due to a difference in quantum efficiency
for the CCDs in the two systems.
Based on the differences in the throughput we expect all the filters at GMOS-N to 
show color terms in the transformations to the SDSS photometric system.

\begin{deluxetable}{lrrr}
\tablecaption{GMOS-N Photometric Zero Points for $u'$\label{tab-photzerou} }
\tablewidth{0pt}
\tabletypesize{\footnotesize}
\tablehead{
\colhead{UT date} & \colhead{$u'_{\rm zero}$} & \colhead{rms($u'$)} & \colhead{N($u'$)} \\
 (yyyymmdd) & 
}
\startdata
20030202 &   25.422 &  0.072 &    8 \\
20030205 &   25.392 &  0.088 &    3 \\
20030922 &   25.341 &  0.126 &    3 \\
20031029 &   25.451 &  0.114 &    5 \\
20031120 &   25.595 &  0.146 &    6 \\
20031121 &   25.472 &  0.109 &    6 \\
20031220 &   25.270 &  0.063 &    2
\enddata
\end{deluxetable}

\subsection{Landolt standard stars on the SDSS system}

The magnitudes from Landolt (1992), which are in the Johnson-Kron-Cousin system, were
transformed to the SDSS system using the transformations from Smith et al.\ (2002), 
except for the transformation involving the $z'$ magnitudes. For the $z'$ magnitudes 
we used the photometry from Smith et al.\ to derive
\begin{equation}
i'-z' = 0.65 (R-I_{\rm c})-0.17 ~~ {\rm rms}=0.027 ~~ N=89
\end{equation}
for $(R-I_{\rm c})<1.00$, and
\begin{equation}
i'-z' = 0.30 (R-I_{\rm c})+0.18 ~~ {\rm rms}=0.016 ~~ N=3 
\end{equation}
for $(R-I_{\rm c})\ge 1.00$.
We also rederived all of the transformations given in 
Smith et al.\ (2002) in order to determine the rms of the transformations.
The transformations for $g'$, $r'$ using $(B-V)$, $(g'-r')$ and $(r'-z')$ have rms values of
0.015 to 0.035. The transformations for $r'$ using $(V-R)$, and $(r'-i')$ have 
rms values $<0.01$. The rms for the transformation for $(u'-g')$ is 0.074.
Except for the $u'$-band, the rms scatter in the transformations is below
or comparable to other sources of uncertainty in the calibration of
the GMOS-N photometry.
The resulting magnitudes in the SDSS system are distributed with the
Gemini IRAF package (v1.5) in the table {\tt gmos\$calib/nlandolt.fits}
and are therefore not listed in this paper.

Figure \ref{fig-colorcolor} shows the color-color diagrams for
the standard stars from Landolt (1992) and Smith et al.\ (2002).
The stars observed with GMOS-N are overplotted.
Four of the standard stars from Landolt (1992) are located off the stellar
sequence in the $(u'-g')$ versus $(g'-r')$ diagram. It is not clear if the
transformations result in incorrect SDSS magnitudes for these stars.
However, the four stars were excluded from the determination of the nightly magnitude
zero points. The stars are SA98\_L5, SA98\_646, SA110\_L1, and SA110\_362.
The very blue object PG0231+051 was also excluded.
The star SA98\_614 was excluded from further analysis because it has a
faint neighbor inside the aperture used for the photometry.
Further, only stars with $0.6\le (r'-z')\le 1.65$ were
included in the determination of the nightly magnitude zero points
for $g'$, $r'$, $i'$, and $z'$, while for $u'$ the stars were also
required to have $(u'-g')\le 2.2$.
This is done to limit the effect of the color terms on the zero points,
see Section 5. For objects with colors outside these ranges it is esseential
that the standard calibration takes into account the color terms derived in Section 5.

While a detailed discussion of the location of various types of stars in the
color-color diagrams is beyond the scope of the present paper, we note
that the standard stars included in Figure 2 separate in two sequences for
colors redder than $(g'-r')=1.2$. The separation is clearest on the $(g'-r')$
versus $(r'-i')$ diagram.  Information from Fan (1999) and Adelman-McCarthy et al.\
(2008) shows that the ``lower'' of the two sequences on this diagram is formed by
low mass main sequence stars, while the ``upper'' sequence contains cool subdwarfs
and possibly carbon stars. We refer to Adelman-McCarthy et al.\ for details on this
as well as details on how to select dwarfs versus giant stars using the SDSS
colors.

\begin{figure*}
\epsfxsize=16.5cm
\epsfbox{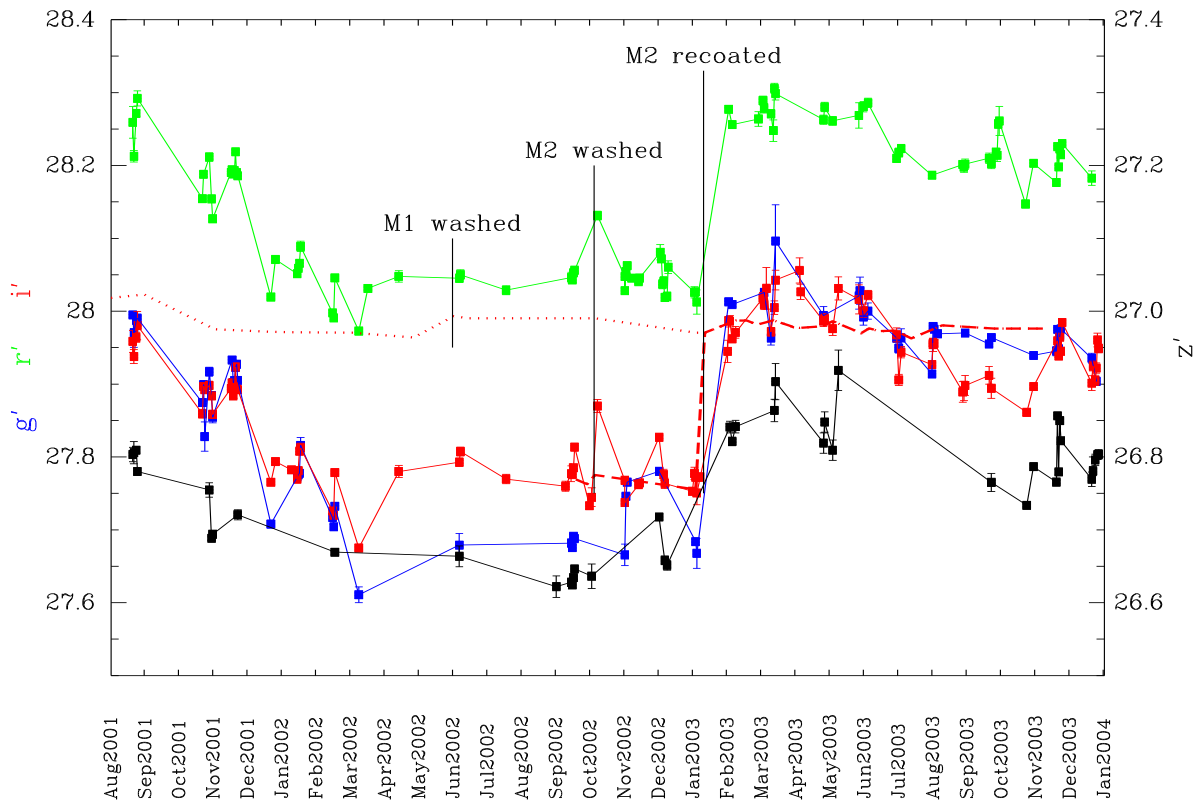}
\caption[]{Magnitude zero points for GMOS-N as a function of UT date.
Solid boxes connected by thin solid lines -- Magnitude zero points listed in Table \ref{tab-photzero}
(blue -- $g'$; green -- $r'$; red -- $i'$; black -- $z'$).
Thick dashed line (red) -- Telescope mirror (primary M1 and secondary M2) reflectivity at 880\,nm 
on a magnitude scale normalized to the magnitude zero point in the $i'$-filter for UT 2003 February 2.
Thick dotted line (red) -- Telescope primary mirror reflectivity at 880\,nm 
on a magnitude scale normalized to the magnitude zero point in the $i'$-filter for UT 2003 February 2.
The washing of the two mirrors as well as the recoating of the secondary mirror are marked.
\label{fig-gmoszero}}
\end{figure*}

\section{Magnitude zero points for GMOS-N \label{sec-mzero}}

\subsection{Nightly magnitude zero points}

For each of the nights and each of the filters, the magnitude zero point 
was derived as the median of $m_{\rm std} - m_{\rm inst} + 28$~~
for each of the unsaturated standard star observations obtained on that night. The resulting magnitude 
zero points, rms and number of standard star measurements used are listed in Tables \ref{tab-photzero}
and \ref{tab-photzerou}.
No color terms were included in this determination.
Using these zero points, approximate standard magnitudes for observations on each of 
these nights may be determined as
\begin{equation}
m_{\rm std} = m_{\rm zero} - 2.5 \log (N/t) - k\,(airmass-1)
\end{equation}
where $t$ is the exposure time, $N$ is the number of electrons
inside the aperture and above the sky level, $airmass$ is the mean airmass for the exposure, 
and $k$ is the median atmospheric extinction at Mauna Kea. 
In Tables \ref{tab-photzero} and \ref{tab-photzerou} the rms scatter of the individual measurements
is listed in the columns ``rms''. Since each zero point is typically based on 7 or more stars
the random uncertainty on the nightly zero points, ${\rm rms}/\sqrt{\rm N}$ where N is the number
of measurements, is typically 0.02 mag or better for $g'$, $r'$, $i'$, and $z'$, and 0.05 mag or
better for $u'$.
However, see Sections 5 and 6 for discussions of color terms and systematic effects, respectively.

\subsection{Magnitude zero point changes with time}

Figure \ref{fig-gmoszero} shows the zero points as a function of the UT date of the night. 
Very large variations in the zero points are seen over periods of weeks to months.
Overplotted on the same figure are the available measurements of the mirror reflectivity 
of Gemini North at 880\,nm.
To make the comparisons easier, these measurements have been converted to magnitudes and normalized to 
the magnitude zero point in the $i'$-filter for UT 2003 February 2.
Reflectivity measurements are also available at 470\,nm, 530\,nm, and 650\,nm.
These show qualitatively the same behavior as the data for 880\,nm and are 
therefore not included on the figure.
During the period covered by Figure \ref{fig-gmoszero} the primary mirror was washed
on UT 2002 May 30. This improved the reflectivity with about 2 per cent.
The secondary mirror was washed UT 2002 October 4, and recoated UT 2003 January 10.
The wash improved the reflectivity with between 2 and 5 per cent, most
in the blue. Prior to the recoating the secondary mirror reflectivity
was between 70 and 80 per cent. The recoating restored the reflectivity to about 90 per cent.
The mirror washes and the secondary mirror recoating are marked on Figure \ref{fig-gmoszero}.

The large decline in system throughput from August 2001 to February 2002 is most likely due to a degradation
of the coating of the secondary mirror. While no reflectivity measurements for the secondary mirror are
available from this period, it is clear that the primary mirror reflectivity did not decline sufficiently
to explain this decline in system throughput.
Further, the system throughput was restored to the original values after the recoating of the secondary
mirror on UT 2003 January 10.

\begin{figure*}
\epsfxsize=16.5cm
\epsfbox{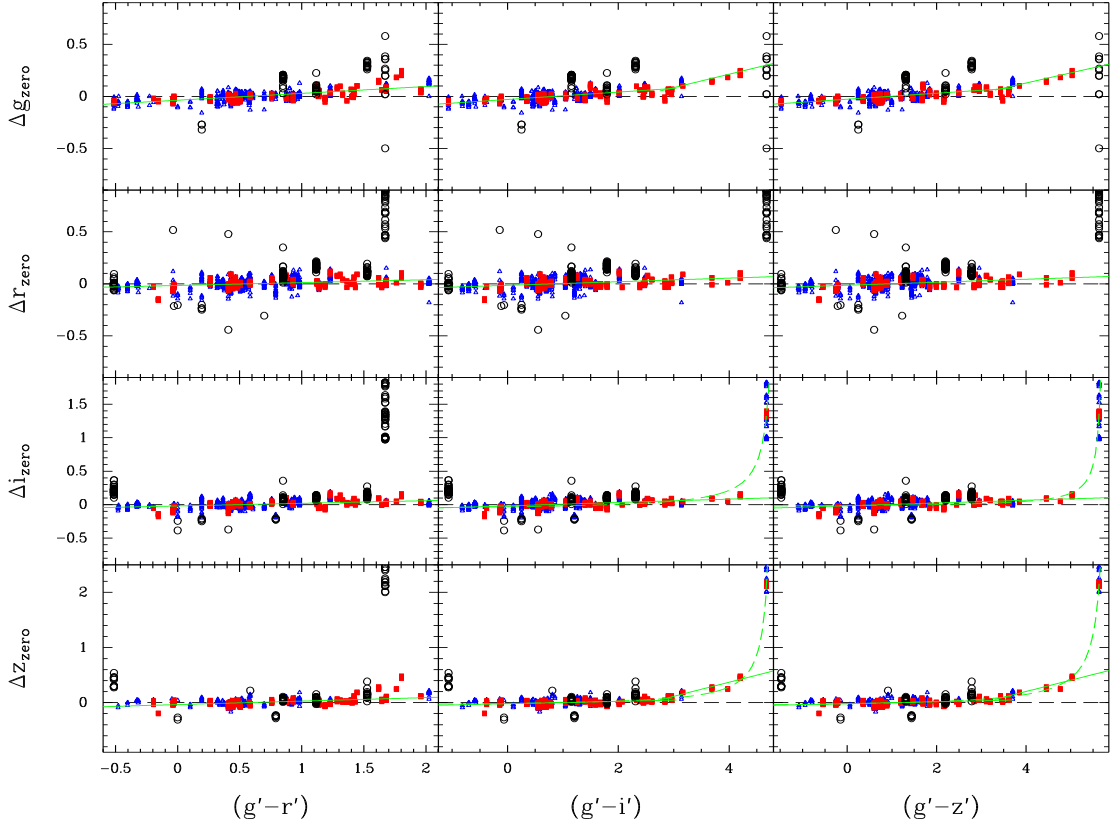}
\caption[]{The residual zero points as a function of the standard colors. Triangles (blue) -- observations obtained
in the period August 2001 to October 2003 and December 2003; boxes (red) -- observations obtained in November 2003;
circles (black) -- measurements not included in the fits, see text for details.
Solid lines (green) -- linear relations listed in Table \ref{tab-color}; 
dashed lines (green) -- inverse color relations listed in Table \ref{tab-color2}.
\label{fig-color1}}
\end{figure*}

Smaller variations in the zero points are most likely due to night-to-night variations in the 
atmospheric extinction.  To quantify these variations, we fitted linear functions to 
the nightly zero points for the $r'$ filter as a function of the number of days after UT 2001 August 1.
The data were fitted in the time intervals 
UT 2001 August 20 to 2002 January 17, UT 2002 February 15 to 2003 January 7, and UT 2003 February 1 
to 2003 December 26.
The rms scatter of the magnitude zero points relative to these fits is $\approx 0.03$ mag.

\newpage
\section{Color terms \label{sec-colorterms}}

For the individual magnitude measurements, the residual zero point was derived as
\begin{equation}
\Delta m_{\rm zero} = m_{\rm std} - m_{\rm inst} + 28 - m_{\rm zero}
\end{equation}
where $m_{\rm zero}$ is the adopted median magnitude zero point for 
the relevant filter on the relevant night. 
Using these residual zero points we can now establish the color terms
using the full database of standard star measurements together.
The $u'$-filter is not included in this analysis, since there are too few 
measurements in the $u'$-filter to reliably establish the color terms.

\begin{figure*}
\epsfxsize=16.5cm
\epsfbox{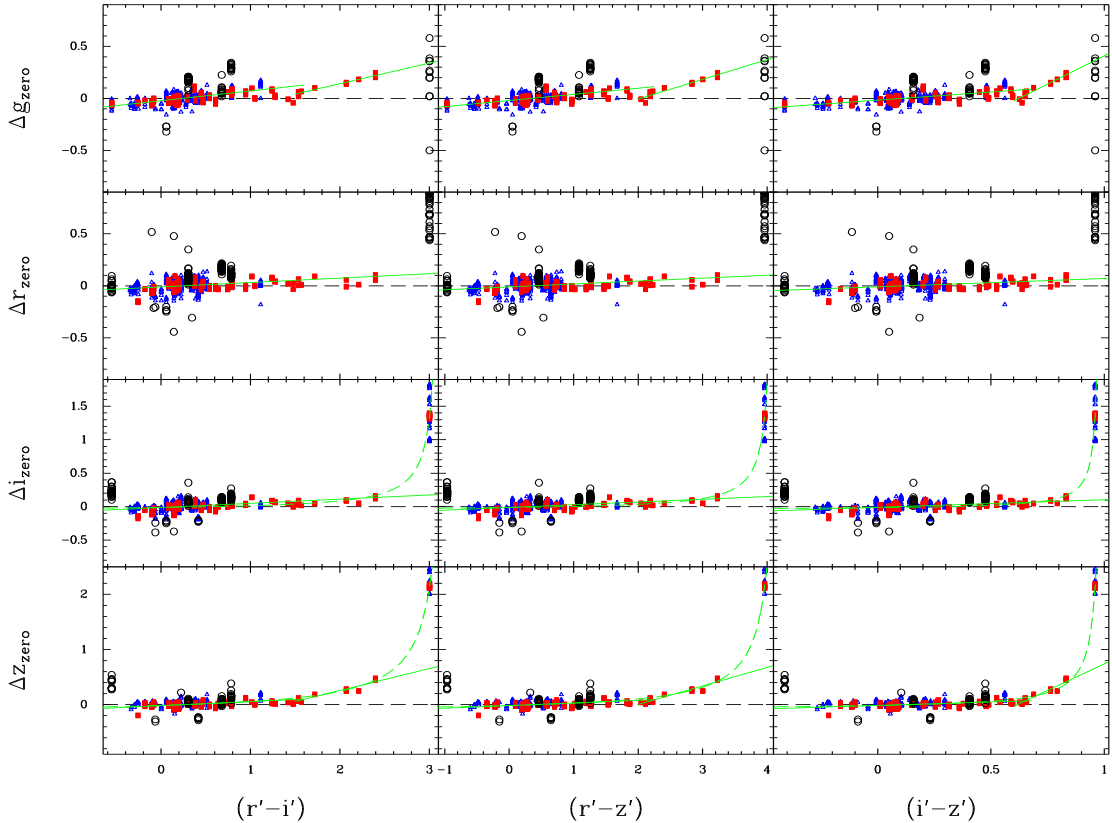}
\caption[]{The residual zero points as a function of the standard colors. Symbols as on Figure \ref{fig-color1}.
\label{fig-color2}}
\end{figure*}

Figures \ref{fig-color1} and \ref{fig-color2} show $\Delta m_{\rm zero}$ 
versus the standard colors of the observed standard stars.  
Table \ref{tab-color} lists the rms scatter of $\Delta m_{\rm zero}$.
This scatter is equivalent to the expected uncertainty on the standard magnitudes if the 
color terms in the calibration are ignored. We fit linear relations to $\Delta m_{\rm zero}$
as a function of the six different colors. In some cases a single linear relation does
not fit the full color range, and we therefore fit two relations, one on the blue and one in the red. 
This is the case for $\Delta g_{\rm zero}$ and $\Delta z_{\rm zero}$ versus all colors except 
the color $(g'-r')$. 
The linear fits and the rms relative to the fits are listed in Table \ref{tab-color}.
The stars excluded from the zero point determinations, were also excluded
from the linear fits, except it turned out that PG0231+051 can be included
in the fits for $\Delta g_{\rm zero}$ without any significant change to the resulting relation.
Further, we iteratively excluded measurements deviating more than 0.2 mag from the relations.
The relations are shown on Figures \ref{fig-color1} and \ref{fig-color2}.

Because many programs executed with GMOS-N target high redshift galaxies with
very red apparent colors, we also established a parameterization for $i'$ and $z'$ that includes
the reddest standard star observed, 98\_L5. In order to do so, we fit relations of the form
\begin{equation}
\label{eq-color2}
\Delta m_{\rm zero} = \alpha + \frac{\beta}{\gamma - C}
\end{equation}
where $C$ is one of the colors. 
The relations $\Delta i_{\rm zero}$ and $\Delta z_{\rm zero}$ versus color $(g'-r')$ 
cannot be fit with this type of function.
The resulting relations should only be seen as parameterization
of the very strong color terms for the $i'$- and $z'$-filters in the far red, and is not
recommended for calibration of observations that do not include very red objects.
The coefficients $\alpha$, $\beta$, and $\gamma$ are listed in Table \ref{tab-color2}.
The relations are shown on Figures \ref{fig-color1} and \ref{fig-color2}.
Because our data do not sample the color intervals from the red end of the linear
relations given in Table \ref{tab-color} and the colors of the star 98\_L5 the values
of $\alpha$ and $\beta$ may be different from the values given here if they were
determined from a dataset sampling these color ranges. However, no such data set
exists for GMOS-N for the time during which Gemini North had an aluminum coating.

Using both the nightly magnitude zero points and the color terms, the standard
magnitudes may be derived as
\begin{equation}
m_{\rm std} = m_{\rm zero} + \Delta m_{\rm zero} - 2.5 \log (N/t) - k\,(airmass-1)
\end{equation}
where $\Delta m_{\rm zero}$ is derived using one of the relations listed in either
Table \ref{tab-color} or Table \ref{tab-color2}.

Since the system throughput shows large variations that all most likely are due to
changes in the telescope mirror coatings, cf.\ Figure \ref{fig-gmoszero} and Section 4.2,
we refitted the color terms in the three time intervals
UT 2001 August 20 to 2002 January 17, UT 2002 February 15 to 2003 January 7, and UT 2003 February 1 to 
2003 December 26.
No significant differences were found between the color terms in the three time intervals,
or between those and the color terms derived using the full sample.
Thus, we conclude that the relations listed in Table \ref{tab-color} and Table \ref{tab-color2}
are valid for the full period from UT 2001 August 20 to UT 2003 December 26.
We also expect the relations to be valid for the priod UT 2003 December 27 to UT 2004 November 21, at
which time the Gemini North primary mirror was recoated with silver instead of aluminum.
This recoating may have lead to a change in the color terms.

\begin{deluxetable*}{llrlrrr}
\tablecaption{GMOS-N Color Terms: Linear Relations \label{tab-color} }
\tablewidth{0pt}
\tablehead{
\colhead{No.} & \colhead{$\Delta {\rm m}$} & \colhead{rms} & \colhead{Color term fit} & \colhead{rms(fit)} & \colhead{N} & Color interval \\
\colhead{(1)} & \colhead{(2)} & \colhead{(3)} & \colhead{(4)} & \colhead{(5)} & \colhead{(6)} & \colhead{(7)}
}
\startdata
1 & $\Delta u_{\rm zero}$   & 0.087\tablenotemark{a} & \nodata & \nodata & 31 & \nodata \\

2 & $\Delta g_{\rm zero}$   & 0.044\tablenotemark{b} & $(0.066\pm 0.002) (g'-r') - (0.037\pm 0.002)$ & 0.034 & 794 & $-0.55\le (g'-r')\le 2.05$\\
3 &       &       & $(0.035\pm 0.002) (g'-i') - (0.029\pm 0.002)$ & 0.033 & 772 & $-0.85\le (g'-i')\le 3.05$ \\
4 &       &       & $(0.129\pm 0.024) (g'-i') - (0.303\pm 0.076)$ & 0.045 & 31 & $2.75\le (g'-i')\le 4.1$ \\
5 &       &       & $(0.029\pm 0.001) (g'-z') - (0.027\pm 0.002)$ & 0.033 & 770 & $-1.1\le (g'-z')\le 3.7$ \\
6 &       &       & $(0.097\pm 0.021) (g'-z') - (0.261\pm 0.079)$ & 0.048 & 31 & $3.3\le (g'-z')\le 5.1$ \\
7 &       &       & $(0.094\pm 0.004) (r'-i') - (0.022\pm 0.002)$ & 0.036 & 788 & $-0.35\le (r'-i')\le 1.6$ \\
8 &       &       & $(0.198\pm 0.019) (r'-i') - (0.255\pm 0.037)$ & 0.017 & 9 & $1.5\le (r'-i')\le 2.4$ \\
9 &       &       & $(0.059\pm 0.002) (r'-z') - (0.020\pm 0.002)$ & 0.035 & 788 & $-0.65\le (r'-z')\le 2.25$ \\
10 &       &       & $(0.189\pm 0.020) (r'-z') - (0.384\pm 0.052)$ & 0.026 & 11 & $2.0\le (r'-z')\le 3.25$ \\
11 &       &       & $(0.156\pm 0.006) (i'-z') - (0.017\pm 0.002)$ & 0.035 & 788 & $-0.3\le (i'-z')\le 0.68$ \\
12 &       &       & $(1.084\pm 0.113) (i'-z') - (0.676\pm 0.082)$ & 0.026 & 11 & $0.6\le (i'-z')\le 0.85$ \\
13 &$\Delta r_{\rm zero}$   & 0.045\tablenotemark{c} & $(0.027\pm 0.003) (g'-r') - (0.016\pm 0.002)$ & 0.043 & 1084 & $-0.55\le (g'-r')\le 2.05$\\
14 &       &       & $(0.017\pm 0.002) (g'-i') - (0.015\pm 0.002)$ & 0.042 & 1084 & $-0.85\le (g'-i')\le 4.1$ \\
15 &       &       & $(0.014\pm 0.001) (g'-z') - (0.014\pm 0.002)$ & 0.042 & 1084 & $-1.1\le (g'-z')\le 5.1$ \\
16 &       &       & $(0.042\pm 0.004) (r'-i') - (0.011\pm 0.002)$ & 0.043 & 1084 & $-0.35\le (r'-i')\le 2.4$ \\
17 &       &       & $(0.028\pm 0.003) (r'-z') - (0.010\pm 0.002)$ & 0.042 & 1084 & $-1.0\le (r'-z')\le 3.25$ \\
18 &       &       & $(0.079\pm 0.007) (i'-z') - (0.009\pm 0.002)$ & 0.042 & 1084 & $-0.45\le (i'-z')\le 0.85$ \\
19 &$\Delta i_{\rm zero}$   & 0.054\tablenotemark{c} & $(0.039\pm 0.003) (g'-r') - (0.020\pm 0.002)$ & 0.050 & 1081 & $-0.55\le (g'-r')\le 2.05$\\
20 &       &       & $(0.025\pm 0.002) (g'-i') - (0.018\pm 0.002)$ & 0.050 & 1081 & $-0.85\le (g'-i')\le 4.1$ \\
21 &       &       & $(0.021\pm 0.002) (g'-z') - (0.017\pm 0.002)$ & 0.050 & 1081 & $-1.1\le (g'-z')\le 5.1$ \\
22 &       &       & $(0.063\pm 0.005) (r'-i') - (0.013\pm 0.002)$ & 0.050 & 1081 & $-0.35\le (r'-i')\le 2.4$ \\
23 &       &       & $(0.041\pm 0.003) (r'-z') - (0.012\pm 0.002)$ & 0.049 & 1081 & $-0.65\le (r'-z')\le 3.25$ \\
24 &       &       & $(0.113\pm 0.008) (i'-z') - (0.010\pm 0.002)$ & 0.049 & 1081 & $-0.3\le (i'-z')\le 0.85$ \\
25 &$\Delta z_{\rm zero}$   & 0.055\tablenotemark{d} & $(0.063\pm 0.005) (g'-r') - (0.035\pm 0.004)$ & 0.057\tablenotemark{e} & 498 & $-0.55\le (g'-r')\le 2.05$\\
26 &       &       & $(0.022\pm 0.003) (g'-i') - (0.019\pm 0.004)$ & 0.046 & 480 & $-0.85\le (g'-i')\le 3.05$ \\
27 &       &       & $(0.256\pm 0.026) (g'-i') - (0.651\pm 0.083)$ & 0.048 & 26 & $2.75\le (g'-i')\le 4.1$ \\
28 &       &       & $(0.018\pm 0.002) (g'-z') - (0.018\pm 0.002)$ & 0.033 & 478 & $-1.1\le (g'-z')\le 3.7$ \\
29 &       &       & $(0.205\pm 0.024) (g'-z') - (0.615\pm 0.094)$ & 0.048 & 26 & $3.3\le (g'-z')\le 5.1$ \\
30 &       &       & $(0.075\pm 0.007) (r'-i') - (0.018\pm 0.003)$ & 0.049 & 492 & $-0.35\le (r'-i')\le 1.6$ \\
31 &       &       & $(0.398\pm 0.053) (r'-i') - (0.539\pm 0.104)$ & 0.047 & 9 & $1.5\le (r'-i')\le 2.4$ \\
32 &       &       & $(0.047\pm 0.005) (r'-z') - (0.017\pm 0.003)$ & 0.050 & 492 & $-0.65\le (r'-z')\le 2.25$ \\
33 &       &       & $(0.336\pm 0.031) (r'-z') - (0.669\pm 0.080)$ & 0.043 & 12 & $2.0\le (r'-z')\le 3.25$ \\
34 &       &       & $(0.125\pm 0.012) (i'-z') - (0.014\pm 0.003)$ & 0.050 & 492 & $-0.3\le (i'-z')\le 0.68$ \\
35 &       &       & $(1.929\pm 0.177) (i'-z') - (1.188\pm 0.127)$ & 0.043 & 12 & $0.6\le (i'-z')\le 0.85$

\enddata
\tablenotetext{a}{$0.2\le (u'-g')\le 2.1$}
\tablenotetext{b}{$-1.1\le (g'-i')\le 3.05$}
\tablenotetext{c}{$-0.7\le (r'-i')\le 2.5$}
\tablenotetext{d}{$-0.7\le (r'-z')\le 2.25$}
\tablenotetext{e}{The rms is higher than without a color term, due to the inclusion of red stars.}
\tablecomments{(1) Calibration number, (2) Residual zero point, (3) rms of $\Delta m$, equivalent to the 
expected uncertainty on the standard calibration if the color terms are ingored, (4) linear fits to the color terms,
(5) rms of the linear fits, (6) number of individual measurements included in the fits, (7) color interval
within which the linear fit applies.}
\end{deluxetable*}

\begin{deluxetable*}{llrrrrrrr}
\tablecaption{GMOS-N Color Terms: Inverse Color Relations \label{tab-color2} }
\tablewidth{0pt}
\tablehead{
\colhead{No.} & \colhead{$\Delta {\rm m}$} & \colhead{Color} & \colhead{$\alpha$} & \colhead{$\beta$} & \colhead{$\gamma$} & \colhead{rms(fit)} & \colhead{N} & Color interval \\
\colhead{(1)} & \colhead{(2)} & \colhead{(3)} & \colhead{(4)} & \colhead{(5)} & \colhead{(6)} & \colhead{(7)} & \colhead{(8)} & \colhead{(9)} }
\startdata
1 & $\Delta i_{\rm zero}$ & $(g'-i')$ & $-0.085\pm0.015$ & $0.255\pm0.038$ & $4.851\pm0.024$ & 0.051 & 1089 & $-0.85\le (g'-i')\le 4.7$ \\
2 & & $(g'-z')$ & $-0.041\pm0.012$ & $0.159\pm0.034$ & $5.748\pm0.023$ & 0.051 & 1089 & $-1.1\le (g'-z')\le 5.65$ \\
3 & & $(r'-i')$ & $-0.067\pm0.016$ & $0.172\pm0.035$ & $3.123\pm0.023$ & 0.051 & 1089 & $-0.35\le (r'-i')\le 3.0$ \\
4 & & $(r'-z')$ & $-0.063\pm0.015$ & $0.204\pm0.040$ & $4.106\pm0.026$ & 0.051 & 1089 & $-0.65\le (r'-z')\le 4.0$ \\
5 & & $(i'-z')$ & $-0.048\pm0.012$ & $0.033\pm0.006$ & $0.983\pm0.004$ & 0.050 & 1089 & $-0.30\le (i'-z')\le 0.95$ \\
6 & $\Delta z_{\rm zero}$ & $(g'-i')$ & $-0.073\pm0.005$ & $0.289\pm0.014$ & $4.804\pm0.006$ & 0.046 & 504 & $-0.85\le (g'-i')\le 4.7$ \\
7 & & $(g'-z')$ &  $-0.097\pm0.005$ & $0.453\pm0.018$ & $5.836\pm0.008$ & 0.046 & 504 & $-1.1\le (g'-z')\le 5.65$ \\
8 & & $(r'-i')$ &  $-0.171\pm0.009$ & $0.505\pm0.025$ & $3.220\pm0.011$ & 0.049 & 504 & $-0.35\le (r'-i')\le 3.0$ \\
9 & & $(r'-z')$ &  $-0.148\pm0.008$ & $0.559\pm0.027$ & $4.206\pm0.012$ & 0.049 & 504 & $-0.65\le (r'-z')\le 4.0$ \\
10 & &$(i'-z')$ &  $-0.091\pm0.006$ & $0.076\pm0.003$ & $0.994\pm0.002$ & 0.049 & 504 & $-0.30\le (i'-z')\le 0.95$

\enddata
\tablecomments{(1) Calibration number. (2) Residual zero point. (3) Color, $C$, in Eq.\ \ref{eq-color2}. 
(4), (5) and (6) Coefficients for the fit, see Eq.\ \ref{eq-color2}. 
(7) rms of the fits. (8) Number of individual measurements included in the fits. (9) Color interval
within which the linear fit applies.}
\end{deluxetable*}

\section{Tests for other systematic effects \label{sec-tests}}

\subsection{Position on the detector array}

The GMOS-N detector array consists of three CCDs with different gains and different 
quantum efficiency.
It is therefore a concern that the three CCDs have been calibrated to consistency.
The large database of standard star observations allow us to test this, as well as test
for other dependences on the position on the detector array.

\begin{figure*}
\epsfxsize=16.5cm
\epsfbox{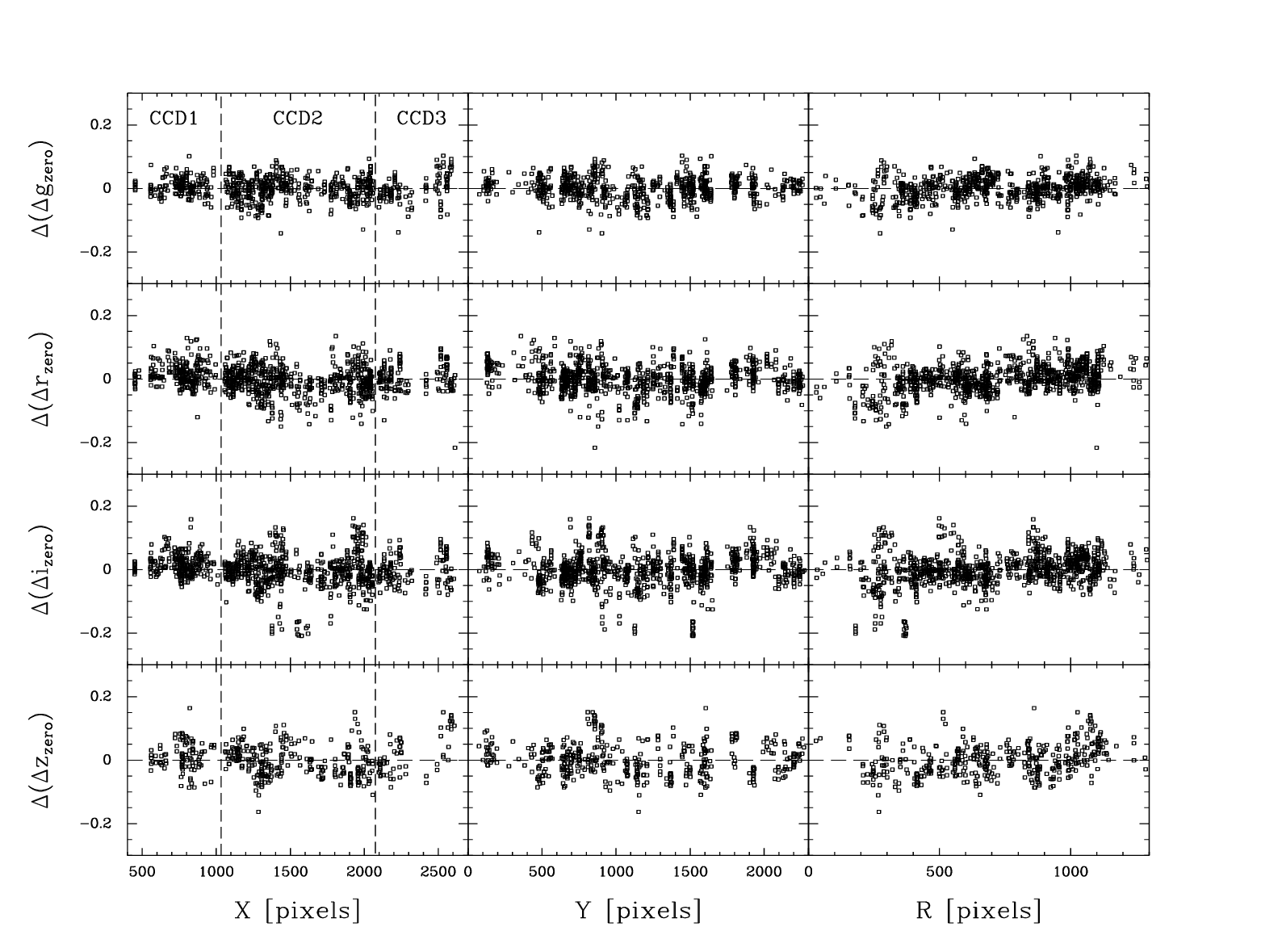}
\caption[]{The color corrected residual zero points, $\Delta (\Delta m_{\rm zero})$,
as a function of the position on the detector array. 
The boundaries between the three detectors in X pixels are shown. 
The GMOS-N imaging field is approximately square
and centered on the rectangular detector array. Thus, the lower limit of the imaging field is around X=400
and the upper limit around X=2700.
\label{fig-dzeroccd}}
\end{figure*}

Figure \ref{fig-dzeroccd} shows the color corrected residual zero points, $\Delta (\Delta m_{\rm zero} )$,
versus the pixel position of the observed standard stars in X and Y 
as well as the distance from the center of the array in pixels.
The color corrected residual zero points are the zero points corrected for the color terms using
the linear relations for $(r'-i')$ from Table \ref{tab-color}. 
Only the stars included in the determination of these relations are included on the figure.
All pixel positions have been transformed to pixels with the detector binned 2\,pixels by 2\,pixels 
as it most often is used for imaging observations. 
With this binning the pixel scale is 0.1454 arcsec per binned pixel.

Table \ref{tab-resmzero} summarizes the mean values of $\Delta (\Delta m_{\rm zero} )$ for the three detectors. 
Detector number one (CCD1) has zero points that are systematically offset from the 
other detectors with about 0.014 mag.
The rms of the residual zero points within a detector is of the same size as the rms for all the measurements.
Thus, the offset of the CCD1 zero points does not significantly
affect the scatter. It is also well within the uncertainty of the standard calibrations taken as part of
the routine calibrations done for GMOS-N photometric data.
We conclude that the individual detectors are calibrated to about 1.5 per cent relative to each other.
The residual color corrected zero points show no significant dependency on the Y pixel coordinate.

\begin{deluxetable}{lrrrrr}
\tablecaption{Mean Color Corrected Residual Zero Points \label{tab-resmzero}}
\tablewidth{0pt}
\tablehead{\colhead{Detector} & \colhead{$g'$} & \colhead{$r'$} & \colhead{$i'$} & \colhead{$z'$} & \colhead{All filters} }
\startdata
CCD1 &  0.007 &  0.015 &  0.018 &  0.012 &  0.014 \\
CCD2 &  0.001 & -0.006 & -0.005 & -0.004 & -0.004 \\
CCD3 & -0.001 &  0.007 & -0.002 &  0.015\tablenotemark{a} & 0.003
\enddata
\tablenotetext{a}{Only 52 measurements.}
\end{deluxetable}

\begin{figure}
\epsfxsize=7cm
\epsfbox{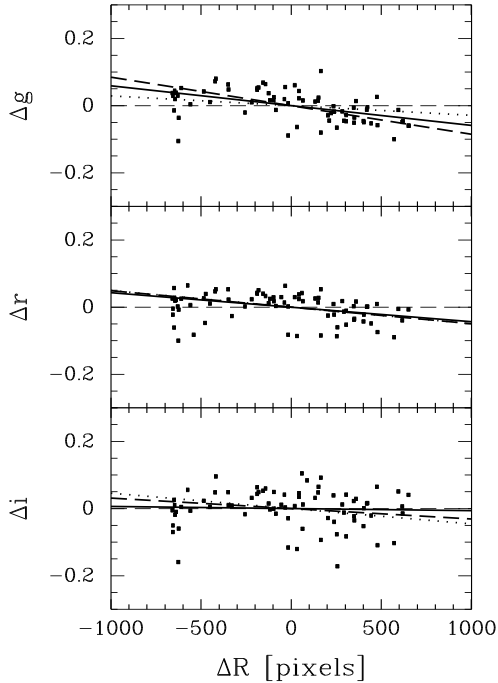}
\caption[]{Sky concentration. Solid boxes -- photometry from the two overlapping fields, the
figure shows the difference in magnitudes versus the difference in the distance from the center
of the field. 
Thick solid line -- best fitting linear relations for the photometry from the two
overlapping fields; 
thick dotted line -- linear relations equivalent to the result based on the standard star photometry;
thick dashed line -- linear relations equivalent to the result based on the
normalized twilight flat fields.
For $\Delta r$ (the $r'$ filter) the three lines are on top of each other.
\label{fig-skyconc}}
\end{figure}

GMOS-N (and GMOS-S at Gemini South) were designed to minimize the effects of sky concentration,
also called parasitic light (Murowinski et al.\ 2003).
Sky concentration originates from light reflected back from the detector to the optics and being
scattered. The scattered light causes an increase in the background level in the center
of the field.
Sky concentration for focal reducer instruments
is discussed in detail by Andersen, Freyhammer \& Storm (1995).
In the presence of sky concentration, we expect the flat fields
to have relatively too high counts in an area around the center of the optical axis.
This in turn will lead to underestimated count levels for stars near the
center of the field.
Andersen et al.\ describe the technique for deriving a 2-dimensional map of the
sky concentration by using offset and rotated images of the same stellar field.
No observations obtained with GMOS-N were done with this purpose in mind.
However, we have used three different methods to estimate the effect
of sky concentration in GMOS-N. (1) The standard star photometry, (2)
a set of observations of two partly overlapping fields observed in the filters
$g'$, $r'$ and $i'$, and (3) normalized twilight flat fields.

If sky concentration affects the standard star photometry, then the 
residual color corrected zero points $\Delta (\Delta m_{\rm zero} )$ are expected to be 
systematically smaller at the center of the field compared to the edges of the field.
Spearman rank order tests were used to test for correlations between 
$\Delta (\Delta m_{\rm zero} )$ and the distance from the field center, $R$, see
Figure \ref{fig-dzeroccd}.
We find significant correlations for all filters and fit linear relations in order
to quantify the size of the effect. Table \ref{tab-skyconc} lists
the effect at the center of the field and 2.5\,arcmin from the center.
Because of the concern that some of this effect may be caused by the zero point offset
for CCD1, we repeated the tests omitting all measurements from CCD1. The correlations
are still present and the effects are of the same size as for the full sample of measurements.

For the observations of the two partly overlapping fields,
we expect the difference in magnitudes for objects in the two fields to depend on the 
difference of the distances from the field center, in such a way that the magnitude 
difference $\Delta m$ decreases with the distance difference $\Delta R$,
cf.\ Andersen et al.\ (1995). 
Figure \ref{fig-skyconc} shows $\Delta m$ versus $\Delta R$ for the three filters.
Using Spearman rank order tests, we find significant correlations for $g'$ and $r'$,
the probabilities that there are no correlations are $\le 0.2$. For $i'$ the test 
shows no significant correlation. 
Figure \ref{fig-skyconc} shows linear fits to the data, as well as the linear relations
equivalent to the result based on the standard star photometry.
The magnitude difference resulting from a difference of 2.5\,arcmin in distance
from the center of the field is listed in Table \ref{tab-skyconc}.

\begin{deluxetable}{lrcc}
\tablecaption{Sky Concentration \label{tab-skyconc}}
\tablewidth{0pt}
\tablehead{\colhead{Filter} & \colhead{Standard stars} & \colhead{Overlapping fields} & \colhead{Flat fields} \\
 & \colhead{$\Delta m_{\rm center}$ to $\Delta m_{\rm 2.5arcmin}$} & \colhead{$\Delta m$}  & \colhead{$\Delta m$} \\
 & \colhead{(1)} & \colhead{(2)} & \colhead{(3)} }
\startdata
$u'$ &  \nodata &  \nodata &  0.103 \\
$g'$ &  -0.020 to 0.009 & 0.059 &  0.085 \\
$r'$ &  -0.036 to 0.015 & 0.043 &  0.049 \\
$i'$ &  -0.032 to 0.015 & $<0.01$ &  0.031 \\
$z'$ &  -0.030 to 0.016 & \nodata &  0.020
\enddata
\tablecomments{(1) The effect of the sky concentration in magnitudes at the 
center of the field, $\Delta m_{\rm center}$,
and 2.5 arcmin from the center, $\Delta m_{\rm 2.5arcmin}$; both relative
to the mean of the standard star measurements.
(2) and (3) The efect of the sky concentration as the magnitude difference between
photometry in the center of the field and 2.5 arcmin from the center.}
\end{deluxetable}

Finally, normalized twilight flats may be used to set limits on the 
size of the sky concentration. The level at the center of the flat fields (we used the 
central 30\,arcsec $\times$ 30\,arcsec)
were compared to the levels at the edges of the field at a distance of 2.5\,arcmin from
the center of the field. The results are summarized in magnitudes in Table \ref{tab-skyconc}.
For comparison, the results are also shown on Figure \ref{fig-skyconc}.

The results from the three methods do not agree completely. However, we can
conclude that there is sky concentration in GMOS-N, and that the effect results in a 
systematic difference of 0.03 to 0.05 mag between photometry derived from objects near
the center of the field and objects near the edges of the field, with the objects near
the center of the field appearing fainter than those near the edges of the field.
The sky concentration may be stronger in the blue than in the red, though the 
standard star photometry seems to indicate that this is not the case.
More detailed engineering data are needed to fully quantify the effect and to derive
a 2-dimensional map of the sky concentration.

\begin{figure}
\epsfxsize=7cm
\epsfbox{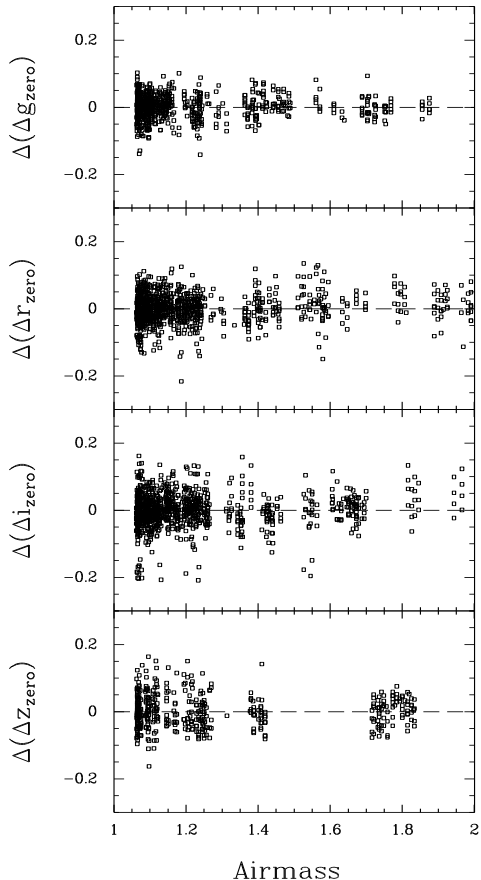}
\caption[]{The color corrected residual zero points, $\Delta (\Delta m_{\rm zero})$, 
as a function of the airmass of the observation.
\label{fig-dzeroair}}
\end{figure}

\subsection{Atmospheric extinction}

The standard stars observed each night cover an airmass interval 
insufficient for determining the nightly atmospheric extinction. 
However, we can perform a simple test of the adopted median atmospheric extinction.
Figure \ref{fig-dzeroair} shows the color corrected residual zero points versus the airmasses. 
For $g'$ and $z'$, the color corrected residual zero points do not correlate with the airmasses. 
Thus, for these filters the adopted median atmospheric extinction is in agreement
with the actual median atmospheric extinction during the nights' of observation.
For $r'$ and $i'$ we find correlations between 
the color corrected residual zero points and the airmasses. 
Spearman rank order tests give probabilities of 0.5 per cent and 0.9 per cent for
$r'$ and $i'$, respectively, that there are no correlations.
If we fit linear relations, then the size of the slopes show that the median 
atmospheric extinction may in fact be about 0.02 smaller during the nights of 
observation than the adopted values.

\section{Summary \label{sec-summary}}

We have determined the nightly magnitude zero points for GMOS-N for the 117 nights in the period
August 2001 to December 2003 during which photometric standard
stars were observed. Long-term variations in the zero points are primarily due to variations in the
reflectivity of the primary and the secondary mirror, with the secondary mirror being the most
important factor.
Night-to-night variations are most likely due to variations in the atmospheric extinction.
All the filters have small color terms for colors in the interval $-0.65\le (r'-z')\le 1.5$.
The maximum effect on the derived standard magnitudes for objects with colors in this interval
is approximately $\pm 0.05$ for $r'$, $i'$ and $z'$, and $\pm 0.07$ for $g'$.
For redder objects the color terms are stronger, especially for the $z'$-filter where
standard magnitudes for objects with $(r'-z')\approx 3.0$ will determined
systematically $\approx 0.35$ mag too bright if the color term is not taken into account.
The color terms presented here are valid for aluminum coating of the primary and secondary
mirror of Gemini North.
The Gemini North primary was recoated with silver during a shutdown in late November 2004.
Thus, we expect the calibrations to be valid for all observations taken in the
period UT 2001 August 20 to UT 2004 November 21, but not for observations obtained at
later dates.

We find that sky concentration affects the photometry with approximately 0.03 to 0.05 mag
over the full field of view. The presence of sky concentration limits the accuracy of photometry
derived from GMOS-N observations using standard methods. Observations specifically designed to map
the sky concentration are needed in order to improve this situation.

\vspace{0.5cm}
The anomymous referee is thanked for useful comments on this paper.
John Thorstensen is thanked for supplying the improved
coordinates for the Landolt standard stars.
Special thanks to the Gemini engineering staff for supplying mirror
reflectivity measurements, and the many student interns at Gemini North
who contributed to the processing of the bias images and flat fields.
Based on observations obtained at the Gemini Observatory, which is operated by the
Association of Universities for Research in Astronomy, Inc., under a cooperative agreement
with the NSF on behalf of the Gemini partnership: the National Science Foundation (United
States), the Science and Technology Facilities Council (United Kingdom), the
National Research Council (Canada), CONICYT (Chile), the Australian Research Council
(Australia), Ministerio da Ciencia e Tecnologia (Brazil) and SECYT (Argentina).
This paper is based on all available standard star observations obtained for all
GMOS-N imaging programs executed on Gemini North in the period from August 2001 to
December 2003.
This research has made use of the USNOFS Image and Catalogue Archive
operated by the United States Naval Observatory, Flagstaff Station
(http://www.nofs.navy.mil/data/fchpix/).

\clearpage

\LongTables
\begin{deluxetable*}{lrrrrrrrrrrrr}
\tablecaption{GMOS-N Photometric Zero Points for $g'\,r'\,i'$ and $z'$\label{tab-photzero} }
\tablewidth{300pt}
\tabletypesize{\footnotesize}
\tablehead{
\colhead{UT date} & \colhead{$g'_{\rm zero}$} & \colhead{rms($g'$)} & \colhead{N($g'$)}  & \colhead{$r'_{\rm zero}$} & \colhead{rms($r'$)} & \colhead{N($r'$)}  & \colhead{$i'_{\rm zero}$} & \colhead{rms($i'$)} & \colhead{N($i'$)}  & \colhead{$z'_{\rm zero}$} & \colhead{rms($z'$)} & \colhead{N($z'$)} \\
 (yyyymmdd) & 
}
\startdata
 20010820& 27.995&  0.005&    2& 28.259&  0.044&    3& 27.959&  0.060&    5& 26.803&  0.040&    5\\
 20010821& 27.970&  \nodata&    1& 28.213&  0.023&    4& 27.938&  0.029&    4& 26.806&  0.046&    4\\
 20010823& 27.988&  0.022&    5& 28.271&  0.037&    9& 27.964&  0.056&    9& 26.809&  0.038&   10\\
 20010824& 27.990&  0.023&    5& 28.292&  0.041&    5& 27.980&  0.059&    4& 26.780&  0.042&   10\\
 20011021& 27.875&  0.038&   20& 28.154&  0.027&    7& 27.859&  0.036&    7&  \nodata&  \nodata&   \nodata\\
 20011022& 27.899&  0.065&   13& 28.188&  0.058&   19& 27.897&  0.073&   25&  \nodata&  \nodata&   \nodata\\
 20011023& 27.828&  \nodata&    1&  \nodata&  \nodata&   \nodata& 27.892&  0.024&    8&  \nodata&  \nodata&   \nodata\\
 20011027& 27.917&  0.012&    3& 28.212&  0.039&    8& 27.898&  0.063&   12& 26.755&  0.030&    4\\
 20011029& 27.884&  0.057&   23& 28.154&  0.071&   29& 27.884&  0.100&   30& 26.688&  0.091&   30\\
 20011030& 27.855&  0.062&    9& 28.127&  0.052&   13& 27.858&  0.066&   13& 26.694&  0.030&   13\\
 20011116&  \nodata&  \nodata&   \nodata& 28.190&  0.034&   28& 27.894&  0.036&   25&  \nodata&  \nodata&   \nodata\\
 20011117& 27.933&  0.016&    7& 28.194&  0.039&   28& 27.902&  0.059&   27&  \nodata&  \nodata&   \nodata\\
 20011118& 27.902&  0.054&   22& 28.188&  0.050&   17& 27.884&  0.049&   15&  \nodata&  \nodata&   \nodata\\
 20011119& 27.905&  0.053&   22& 28.193&  0.043&   16& 27.896&  0.046&   16&  \nodata&  \nodata&   \nodata\\
 20011120& 27.923&  0.045&   21& 28.219&  0.053&   24& 27.923&  0.080&   20&  \nodata&  \nodata&   \nodata\\
 20011121& 27.927&  0.036&   16& 28.190&  0.032&   14& 27.893&  0.043&   14&  \nodata&  \nodata&   \nodata\\
 20011122& 27.905&  0.042&   15& 28.186&  0.054&   16& 27.892&  0.076&   16& 26.721&  0.103&   16\\
 20011221& 27.708&  0.056&   12& 28.019&  0.059&   14& 27.765&  0.057&   14&  \nodata&  \nodata&   \nodata\\
 20011225&  \nodata&  \nodata&   \nodata& 28.071&  0.044&   12& 27.794&  0.046&   12&  \nodata&  \nodata&   \nodata\\
 20020109&  \nodata&  \nodata&   \nodata&  \nodata&  \nodata&   \nodata& 27.782&  0.026&   25&  \nodata&  \nodata&   \nodata\\
 20020114& 27.772&  0.040&   16& 28.051&  0.033&   15& 27.770&  0.039&   13&  \nodata&  \nodata&   \nodata\\
 20020115& 27.775&  0.041&   16& 28.059&  0.037&   15& 27.781&  0.039&   14&  \nodata&  \nodata&   \nodata\\
 20020116& 27.778&  0.028&    8& 28.066&  0.031&    8& 27.808&  0.030&    8&  \nodata&  \nodata&   \nodata\\
 20020117& 27.816&  0.033&    4& 28.089&  0.049&    8& 27.812&  0.045&    8&  \nodata&  \nodata&   \nodata\\
 20020215& 27.717&  0.080&   17& 27.998&  0.068&   24& 27.726&  0.063&   24&  \nodata&  \nodata&   \nodata\\
 20020216& 27.704&  0.058&   20& 27.991&  0.045&   23& 27.720&  0.049&   23&  \nodata&  \nodata&   \nodata\\
 20020217& 27.732&  0.047&   47& 28.046&  0.113&   47& 27.779&  0.044&   49& 26.669&  0.044&   50\\
 20020308& 27.611&  0.043&    5& 27.973&  0.026&    6& 27.675&  0.024&    6&  \nodata&  \nodata&   \nodata\\
 20020316&  \nodata&  \nodata&   \nodata& 28.031&  0.022&    8&  \nodata&  \nodata&   \nodata&  \nodata&  \nodata&   \nodata\\
 20020413&  \nodata&  \nodata&   \nodata& 28.048&  0.039&    6& 27.780&  0.042&    6&  \nodata&  \nodata&   \nodata\\
 20020606& 27.679&  0.080&    6& 28.045&  0.039&    8& 27.793&  0.033&    7& 26.664&  0.086&    7\\
 20020607&  \nodata&  \nodata&   \nodata& 28.050&  0.040&    7& 27.807&  0.037&    7&  \nodata&  \nodata&   \nodata\\
 20020717&  \nodata&  \nodata&   \nodata& 28.029&  0.043&    8& 27.770&  0.046&    8&  \nodata&  \nodata&   \nodata\\
 20020901&  \nodata&  \nodata&   \nodata&  \nodata&  \nodata&   \nodata&  \nodata&  \nodata&   \nodata& 26.622&  0.103&    8\\
 20020909&  \nodata&  \nodata&   \nodata&  \nodata&  \nodata&   \nodata& 27.760&  0.063&   10&  \nodata&  \nodata&   \nodata\\
 20020914& 27.682&  0.038&    8& 28.046&  0.044&    7& 27.777&  0.065&    7& 26.628&  0.040&    8\\
 20020915& 27.676&  0.063&   16& 28.043&  0.043&   14& 27.776&  0.052&   13& 26.624&  0.062&   15\\
 20020916& 27.691&  0.059&   24& 28.053&  0.046&   21& 27.785&  0.064&   21& 26.634&  0.058&   23\\
 20020917& 27.688&  0.034&   16& 28.056&  0.046&   16& 27.814&  0.067&   15& 26.646&  0.038&   16\\
 20020930&  \nodata&  \nodata&   \nodata&  \nodata&  \nodata&   \nodata& 27.733&  0.034&    8&  \nodata&  \nodata&   \nodata\\
 20021002&  \nodata&  \nodata&   \nodata&  \nodata&  \nodata&   \nodata& 27.745&  0.116&   10& 26.636&  0.151&   10\\
 20021007&  \nodata&  \nodata&   \nodata& 28.131&  0.042&    8& 27.870&  0.062&    8&  \nodata&  \nodata&   \nodata\\
 20021031& 27.666&  0.044&    4& 28.048&  0.046&   22& 27.768&  0.050&   12&  \nodata&  \nodata&   \nodata\\
 20021101&  \nodata&  \nodata&   \nodata& 28.028&  0.036&   12& 27.737&  0.040&   12&  \nodata&  \nodata&   \nodata\\
 20021102& 27.746&  0.040&   16& 28.054&  0.032&   14&  \nodata&  \nodata&   \nodata&  \nodata&  \nodata&   \nodata\\
 20021103& 27.765&  0.042&   16& 28.062&  0.031&   14&  \nodata&  \nodata&   \nodata&  \nodata&  \nodata&   \nodata\\
 20021107&  \nodata&  \nodata&   \nodata& 28.045&  0.039&   17&  \nodata&  \nodata&   \nodata&  \nodata&  \nodata&   \nodata\\
 20021113&  \nodata&  \nodata&   \nodata& 28.041&  0.045&   29& 27.762&  0.037&   12&  \nodata&  \nodata&   \nodata\\
 20021114&  \nodata&  \nodata&   \nodata& 28.046&  0.041&   15& 27.764&  0.034&   11&  \nodata&  \nodata&   \nodata\\
 20021201& 27.780&  0.049&   17&  \nodata&  \nodata&   \nodata& 27.827&  0.039&   16& 26.718&  0.044&   17\\
 20021202&  \nodata&  \nodata&   \nodata& 28.081&  0.136&   14&  \nodata&  \nodata&   \nodata&  \nodata&  \nodata&   \nodata\\
 20021203&  \nodata&  \nodata&   \nodata& 28.072&  0.052&   13&  \nodata&  \nodata&   \nodata&  \nodata&  \nodata&   \nodata\\
 20021204&  \nodata&  \nodata&   \nodata& 28.037&  0.044&    8&  \nodata&  \nodata&   \nodata&  \nodata&  \nodata&   \nodata\\
 20021205&  \nodata&  \nodata&   \nodata& 28.041&  0.045&    8& 27.776&  0.052&    8&  \nodata&  \nodata&   \nodata\\
 20021206&  \nodata&  \nodata&   \nodata& 28.019&  0.034&    7& 27.764&  0.049&    8& 26.658&  0.042&    8\\
 20021208&  \nodata&  \nodata&   \nodata& 28.020&  0.043&    8&  \nodata&  \nodata&   \nodata& 26.651&  0.048&    8\\
 20021209&  \nodata&  \nodata&   \nodata& 28.060&  0.060&    8&  \nodata&  \nodata&   \nodata&  \nodata&  \nodata&   \nodata\\
 20021230&  \nodata&  \nodata&   \nodata&  \nodata&  \nodata&   \nodata& 27.753&  0.023&    8&  \nodata&  \nodata&   \nodata\\
 20030102&  \nodata&  \nodata&   \nodata& 28.026&  0.054&    8& 27.777&  0.060&    8&  \nodata&  \nodata&   \nodata\\
 20030103& 27.684&  0.026&    8& 28.027&  0.035&    8& 27.772&  0.037&    8&  \nodata&  \nodata&   \nodata\\
 20030104& 27.668&  0.144&    8& 28.013&  0.117&    8& 27.751&  0.114&    8&  \nodata&  \nodata&   \nodata\\
 20030107&  \nodata&  \nodata&   \nodata&  \nodata&  \nodata&   \nodata& 27.772&  0.040&    8&  \nodata&  \nodata&   \nodata\\
 20030201&  \nodata&  \nodata&   \nodata&  \nodata&  \nodata&   \nodata& 27.945&  0.106&    8&  \nodata&  \nodata&   \nodata\\
 20030202& 28.013&  0.023&    8& 28.277&  0.023&    8& 27.987&  0.031&    8&  \nodata&  \nodata&   \nodata\\
 20030203&  \nodata&  \nodata&   \nodata&  \nodata&  \nodata&   \nodata& 27.986&  0.056&    8& 26.841&  0.056&    8\\
 20030205& 28.009&  0.026&   15& 28.256&  0.027&    8& 27.962&  0.029&    8& 26.821&  0.041&    8\\
 20030208&  \nodata&  \nodata&   \nodata&  \nodata&  \nodata&   \nodata& 27.971&  0.057&    8& 26.842&  0.061&    8\\
 20030228&  \nodata&  \nodata&   \nodata& 28.264&  0.070&    8&  \nodata&  \nodata&   \nodata&  \nodata&  \nodata&   \nodata\\
 20030302&  \nodata&  \nodata&   \nodata& 28.289&  0.043&    8& 28.018&  0.051&    8&  \nodata&  \nodata&   \nodata\\
 20030303& 28.026&  0.035&    8& 28.278&  0.046&    8& 28.009&  0.051&    8&  \nodata&  \nodata&   \nodata\\
 20030305&  \nodata&  \nodata&   \nodata&  \nodata&  \nodata&   \nodata& 28.031&  0.085&    4&  \nodata&  \nodata&   \nodata\\
 20030309& 27.963&  0.048&    6& 28.271&  0.039&    8& 27.972&  0.042&    8&  \nodata&  \nodata&   \nodata\\
 20030311&  \nodata&  \nodata&   \nodata& 28.248&  0.104&    8&  \nodata&  \nodata&   \nodata&  \nodata&  \nodata&   \nodata
\end{deluxetable*}

\LongTables
\begin{deluxetable*}{lrrrrrrrrrrrr}
\tablecaption{GMOS-N Photometric Zero Points for $g'\,r'\,i'$ and $z'$}
\tablewidth{300pt}
\tabletypesize{\footnotesize}
\tablenum{7}
\tablehead{
\colhead{UT date} & \colhead{$g'_{\rm zero}$} & \colhead{rms($g'$)} & \colhead{N($g'$)}  & \colhead{$r'_{\rm zero}$} & \colhead{rms($r'$)} & \colhead{N($r'$)}  & \colhead{$i'_{\rm zero}$} & \colhead{rms($i'$)} & \colhead{N($i'$)}  & \colhead{$z'_{\rm zero}$} & \colhead{rms($z'$)} & \colhead{N($z'$)} \\
 (yyyymmdd) & 
}
\startdata
 20030312&  \nodata&  \nodata&   \nodata& 28.306&  0.034&    6& 28.005&  0.037&    5& 26.864&  0.092&    7\\
 20030313& 28.096&  0.099&    3& 28.299&  0.044&    6& 28.043&  0.054&    5& 26.903&  0.100&    5\\
 20030404&  \nodata&  \nodata&   \nodata&  \nodata&  \nodata&   \nodata& 28.056&  0.069&    5&  \nodata&  \nodata&   \nodata\\
 20030405&  \nodata&  \nodata&   \nodata&  \nodata&  \nodata&   \nodata& 28.027&  0.042&    5&  \nodata&  \nodata&   \nodata\\
 20030425& 27.994&  0.084&    8& 28.263&  0.037&    7& 27.986&  0.036&    6& 26.819&  0.084&    7\\
 20030426&  \nodata&  \nodata&   \nodata& 28.280&  0.045&    8& 27.990&  0.043&    7& 26.848&  0.083&    7\\
 20030503&  \nodata&  \nodata&   \nodata& 28.261&  0.036&    7& 27.976&  0.046&    6& 26.809&  0.084&    7\\
 20030508&  \nodata&  \nodata&   \nodata&  \nodata&  \nodata&   \nodata& 28.031&  0.079&    6& 26.919&  0.139&    6\\
 20030526& 28.021&  0.090&    6& 28.269&  0.052&    4& 28.016&  0.060&    4&  \nodata&  \nodata&   \nodata\\
 20030527& 28.028&  0.093&    6&  \nodata&  \nodata&   \nodata&  \nodata&  \nodata&   \nodata&  \nodata&  \nodata&   \nodata\\
 20030530& 27.992&  0.078&    8& 28.281&  0.042&    7& 28.001&  0.036&    5&  \nodata&  \nodata&   \nodata\\
 20030604& 28.000&  0.079&    8& 28.286&  0.045&    8& 28.022&  0.032&    6&  \nodata&  \nodata&   \nodata\\
 20030629& 27.963&  0.041&    8& 28.210&  0.031&    7& 27.968&  0.058&    7&  \nodata&  \nodata&   \nodata\\
 20030701& 27.949&  0.044&    8& 28.218&  0.037&    7& 27.906&  0.046&    7&  \nodata&  \nodata&   \nodata\\
 20030703& 27.963&  0.037&    4& 28.223&  0.027&    6& 27.944&  0.039&    6&  \nodata&  \nodata&   \nodata\\
 20030730& 27.914&  0.031&    8& 28.187&  0.032&    8& 27.927&  0.035&    8&  \nodata&  \nodata&   \nodata\\
 20030731&  \nodata&  \nodata&   \nodata&  \nodata&  \nodata&   \nodata& 27.954&  0.045&    6&  \nodata&  \nodata&   \nodata\\
 20030801& 27.979&  0.039&   15&  \nodata&  \nodata&   \nodata& 27.956&  0.055&   10&  \nodata&  \nodata&   \nodata\\
 20030802& 27.975&  0.032&   10&  \nodata&  \nodata&   \nodata& 27.956&  0.049&   10&  \nodata&  \nodata&   \nodata\\
 20030805& 27.969&  0.044&   10&  \nodata&  \nodata&   \nodata&  \nodata&  \nodata&   \nodata&  \nodata&  \nodata&   \nodata\\
 20030827&  \nodata&  \nodata&   \nodata& 28.201&  0.050&   15& 27.889&  0.114&    9&  \nodata&  \nodata&   \nodata\\
 20030828&  \nodata&  \nodata&   \nodata& 28.198&  0.099&   15& 27.891&  0.107&    9&  \nodata&  \nodata&   \nodata\\
 20030829& 27.970&  0.023&   10& 28.202&  0.057&    9& 27.898&  0.110&    9&  \nodata&  \nodata&   \nodata\\
 20030920& 27.955&  0.050&   15& 28.210&  0.056&    9& 27.912&  0.108&   10&  \nodata&  \nodata&   \nodata\\
 20030922& 27.964&  0.022&   10& 28.203&  0.058&    9& 27.894&  0.111&    9& 26.765&  0.124&   11\\
 20030926&  \nodata&  \nodata&   \nodata& 28.218&  0.054&   10&  \nodata&  \nodata&   \nodata&  \nodata&  \nodata&   \nodata\\
 20030927&  \nodata&  \nodata&   \nodata& 28.214&  0.068&    9&  \nodata&  \nodata&   \nodata&  \nodata&  \nodata&   \nodata\\
 20030928&  \nodata&  \nodata&   \nodata& 28.257&  0.042&    8&  \nodata&  \nodata&   \nodata&  \nodata&  \nodata&   \nodata\\
 20030929&  \nodata&  \nodata&   \nodata& 28.261&  0.020&    2&  \nodata&  \nodata&   \nodata&  \nodata&  \nodata&   \nodata\\
 20031022&  \nodata&  \nodata&   \nodata& 28.147&  0.033&    7&  \nodata&  \nodata&   \nodata&  \nodata&  \nodata&   \nodata\\
 20031023&  \nodata&  \nodata&   \nodata&  \nodata&  \nodata&   \nodata& 27.861&  0.038&    9& 26.734&  0.042&   10\\
 20031029& 27.939&  0.039&   15& 28.203&  0.043&   12& 27.897&  0.032&    9& 26.787&  0.059&   14\\
 20031119& 27.945&  0.006&    2& 28.177&  0.003&    2&  \nodata&  \nodata& \nodata& 26.765&  0.003&    2\\
 20031120& 27.975&  0.043&   31& 28.226&  0.024&   31& 27.959&  0.078&   31& 26.856&  0.045&   31\\
 20031121& 27.948&  0.034&   17& 28.198&  0.032&   14& 27.939&  0.029&   15& 26.779&  0.051&   17\\
 20031122& 27.967&  0.047&   22& 28.223&  0.046&   21& 27.964&  0.063&   20& 26.850&  0.047&   21\\
 20031123& 27.976&  0.043&   15& 28.215&  0.036&   12& 27.945&  0.040&   11& 26.822&  0.045&   16\\
 20031124&  \nodata&  \nodata&   \nodata& 28.230&  0.039&   15& 27.985&  0.056&   14&  \nodata&  \nodata&   \nodata \\
 20031220& 27.935&  0.034&    6& 28.183&  0.050&    6& 27.901&  0.051&    6& 26.770&  0.049&    6\\
 20031221&  \nodata&  \nodata&   \nodata&  \nodata&  \nodata&   \nodata& 27.925&  0.047&    8& 26.781&  0.044&    8\\
 20031223&  \nodata&  \nodata&   \nodata&  \nodata&  \nodata&   \nodata& 27.905&  0.049&    8& 26.796&  0.056&    8\\
 20031224& 27.904&  0.041&    8&  \nodata&  \nodata&   \nodata& 27.922&  0.049&    8&  \nodata&  \nodata&   \nodata\\
 20031225&  \nodata&  \nodata&   \nodata&  \nodata&  \nodata&   \nodata& 27.960&  0.067&    8& 26.803&  0.043&    8\\
 20031226&  \nodata&  \nodata&   \nodata&  \nodata&  \nodata&   \nodata& 27.949&  0.041&    8& 26.804&  0.045&    8
\enddata
\end{deluxetable*}

\end{document}